\documentclass[11pt,letterpaper]{article}

\usepackage[utf8]{inputenc}
\usepackage[T1]{fontenc}
\usepackage[margin=1in]{geometry}
\usepackage{amsmath,amssymb}
\usepackage{graphicx}
\usepackage{booktabs}
\usepackage{tabularx}
\usepackage{multirow}
\usepackage{enumitem}
\usepackage{xcolor}
\usepackage{float}
\usepackage[breaklinks=true]{hyperref}
\usepackage{url}
\usepackage{cite}
\usepackage{fancyhdr}
\usepackage{titlesec}
\usepackage{setspace}
\usepackage{tcolorbox}

\definecolor{accentblue}{HTML}{0D47A1}

\hypersetup{
  colorlinks=true,
  linkcolor=accentblue,
  citecolor=accentblue,
  urlcolor=accentblue,
  pdfauthor={Saikat Maiti},
  pdftitle={Caging the Agents: A Zero Trust Security Architecture for Autonomous AI in Healthcare}
}

\titleformat{\section}{\large\bfseries}{\thesection}{1em}{}
\titleformat{\subsection}{\normalsize\bfseries}{\thesubsection}{1em}{}
\titleformat{\subsubsection}{\normalsize\itshape}{\thesubsubsection}{1em}{}

\newtcolorbox{keyfindings}{
  colback=blue!3, colframe=accentblue, fonttitle=\bfseries, title=Key Finding,
  boxrule=0.5pt, left=6pt, right=6pt, top=4pt, bottom=4pt
}

\newtcolorbox{threatbox}[1]{
  colback=red!3, colframe=red!60!black, fonttitle=\bfseries, title=#1,
  boxrule=0.5pt, left=6pt, right=6pt, top=4pt, bottom=4pt
}

\begin{document}

\begin{center}
  {\LARGE\bfseries Caging the Agents: A Zero Trust Security\\[4pt]
  Architecture for Autonomous AI in Healthcare}\\[18pt]
  {\large Saikat Maiti}\\[6pt]
  {\normalsize VP of Trust, Commure\\
  Founder \& CEO, nFactor Technologies}\\[6pt]
  {\small\texttt{saikat@nfactor.ai}}\\[12pt]
  {\normalsize March 2026}\\[6pt]
  {\small Version 1.0}
\end{center}

\vspace{12pt}\hrule\vspace{16pt}

\begin{abstract}
\noindent
Autonomous AI agents powered by large language models are being deployed in production environments with capabilities that include shell execution, file system access, database queries, HTTP requests, and multi party communication. Recent empirical research has demonstrated that these agents exhibit critical security vulnerabilities when deployed in realistic settings, including unauthorized compliance with non owner instructions, disclosure of sensitive information, identity spoofing, cross agent propagation of unsafe practices, and susceptibility to indirect prompt injection through external editable resources~\cite{shapira2026agents}. When these agents operate within healthcare infrastructure processing Protected Health Information (PHI), every documented vulnerability becomes a potential HIPAA violation.

This paper presents a comprehensive security architecture developed and deployed for a fleet of nine autonomous AI agents running in production at a healthcare technology company. The architecture addresses the six domain threat model we developed for agentic AI in healthcare: credential exposure, execution capability abuse, network egress exfiltration, prompt integrity failures, database access risks, and fleet configuration drift. We implement a four layer defense in depth approach: (1)~kernel level workload isolation using gVisor sandboxed containers on Kubernetes, (2)~credential proxy sidecars that prevent agent containers from accessing raw secrets, (3)~network egress policies enforced at the Kubernetes NetworkPolicy layer restricting each agent to allowlisted destinations, and (4)~a prompt integrity framework with cryptographically structured metadata envelopes and explicit untrusted content labeling.

We report empirical results from a 90 day deployment, including four HIGH severity findings discovered and remediated by an automated security audit agent, the progressive hardening of the fleet from an unhardened baseline to the target architecture, and the security posture metrics before and after control deployment. We map each documented vulnerability from recent red teaming research to the specific defensive control that addresses it, demonstrating coverage across all eleven attack patterns identified in the literature. All architecture specifications, Kubernetes configurations, audit tooling, and the prompt integrity framework are released as open source.
\end{abstract}


\newpage
\tableofcontents
\newpage

\section{Introduction}
\label{sec:introduction}

The deployment of autonomous AI agents in production environments represents a qualitative shift in the security landscape. Unlike conventional software that processes inputs through well defined interfaces, autonomous agents powered by large language models (LLMs) operate with capabilities that blur the boundary between tool and operator: they execute shell commands, read and write files, query databases, make HTTP requests to external services, spawn sub agents, and maintain persistent memory across sessions~\cite{openclaw, shapira2026agents}. These capabilities, combined with natural language instruction processing from multiple communication channels, create an attack surface that existing security frameworks were not designed to address.

The urgency of this challenge is underscored by recent empirical research. Shapira et al.~\cite{shapira2026agents} conducted a two week red teaming study of autonomous agents deployed in a live laboratory environment using the OpenClaw framework, documenting eleven representative failure modes including unauthorized compliance with non owner instructions, disclosure of 124 email records to an unauthorized party, identity spoofing through display name manipulation, agent corruption via indirect prompt injection through external editable resources, cross agent propagation of unsafe practices, and denial of service through uncontrolled resource consumption. Their findings establish that these are not theoretical risks but empirically demonstrated vulnerabilities in realistic deployment settings.

When autonomous agents with these capabilities operate within healthcare infrastructure, the stakes are fundamentally different. Every vulnerability documented by Shapira et al. maps to a potential HIPAA violation: an agent that discloses email records containing Protected Health Information to an unauthorized party triggers breach notification obligations; an agent that accepts instructions from a spoofed identity may execute operations on clinical data systems; an agent corrupted through indirect prompt injection may exfiltrate patient data to attacker controlled destinations. The NIST AI Agent Standards Initiative, announced in February 2026, identifies agent identity, authorization, and security as priority areas for standardization~\cite{nist_agent_standards}, but provides no implementation guidance for healthcare deployments.

This paper addresses the gap between documented vulnerabilities and deployed defenses. We present the complete security architecture developed for a fleet of nine autonomous AI agents operating in production at a healthcare technology company whose subsidiaries serve major hospital networks including clinical AI, ambient documentation, and patient engagement systems. The fleet uses the OpenClaw framework~\cite{openclaw} with Claude via AWS Bedrock for model inference, deployed on Google Cloud Platform Compute Engine infrastructure.

Our contributions are as follows:

\begin{enumerate}[leftmargin=2em]
  \item A six domain threat model for autonomous AI agents in healthcare that maps every attack vector to specific HIPAA Security Rule provisions, incorporating the empirical findings from Shapira et al.~\cite{shapira2026agents} as validated threat scenarios.
  \item A four layer defense in depth architecture (kernel isolation, credential proxy, network egress policy, prompt integrity framework) designed specifically for agentic AI workloads on Kubernetes with Temporal workflow orchestration.
  \item An automated fleet security audit system (itself an AI agent) that continuously scans for credential exposure, permission drift, and configuration divergence, with empirical results from production operation including four HIGH severity findings discovered and remediated.
  \item A 90 day longitudinal dataset documenting the progressive hardening of a production agentic AI fleet, from unhardened baseline through three hardening generations, with security posture metrics at each stage.
\end{enumerate}

\section{Background and Related Work}
\label{sec:background}

\subsection{Autonomous AI Agent Architectures}

Autonomous AI agents are LLM powered entities that can plan and take actions to execute goals over multiple iterations~\cite{shapira2026agents}. The OpenClaw framework~\cite{openclaw} provides a representative architecture: agents are instantiated as long running services with an owner (a primary human operator), a dedicated machine (a virtual machine with persistent storage), and multiple communication surfaces (messaging platforms and email) through which both owners and non owners can interact with the agent.

OpenClaw agents are configured through markdown files in the agent's workspace directory. The configuration includes persona, operating instructions, tool conventions, and user profile, stored across several workspace files (AGENTS.md, SOUL.md, TOOLS.md, IDENTITY.md, USER.md) that are injected into the model's context on every turn. Critically, all of these files, including the agent's own operating instructions, can be modified by the agent itself, allowing it to update its behavior and memory through conversation. Agents have unrestricted shell access, file system access, package installation capabilities, and the ability to communicate via messaging platforms and email.

\subsection{Documented Vulnerabilities in Agentic AI Systems}

Shapira et al.~\cite{shapira2026agents} provide the most comprehensive empirical documentation of agentic AI vulnerabilities to date. Their eleven case studies reveal three structural deficiencies in current agent architectures that are directly relevant to our security design:

\textbf{No stakeholder model.} Agents lack a coherent representation of who they serve, who they interact with, and what obligations they have to each party. In practice, agents default to satisfying whoever is speaking most urgently, recently, or coercively. This is the most commonly exploited attack surface: agents in their study executed filesystem commands for any non owner who asked, disclosed 124 email records including sensitive information to an unauthorized party, and complied with system shutdown instructions from a spoofed identity~\cite{shapira2026agents}.

\textbf{No self model.} Agents take irreversible, user affecting actions without recognizing they are exceeding their own competence boundaries. Agents converted short lived conversational requests into permanent background processes with no termination condition, allocated memory indefinitely without recognizing operational threats, and reported task completion while the underlying system state contradicted those reports~\cite{shapira2026agents}.

\textbf{Instruction data conflation.} LLM based agents process instructions and data as tokens in a context window, making the two fundamentally indistinguishable. Prompt injection is therefore a structural feature of these systems rather than a fixable bug. Schmotz et al.~\cite{schmotz2025agent_skills} demonstrate that agent skill files (markdown files loaded into context) enable realistic, trivially simple prompt injections that can drive data exfiltration. Zhang et al.~\cite{zhang2025prompt_loops} show that prompt injection can induce infinite action loops with over 80 percent success.

\subsection{Regulatory Context for Healthcare Agentic AI}

The HIPAA Security Rule~\cite{hipaa_security_rule} establishes requirements for the protection of electronic PHI that apply to autonomous agent deployments. Several provisions are particularly relevant: access controls (45~CFR~164.312(a)) requiring that only authorized persons or software programs access ePHI; audit controls (45~CFR~164.312(b)) requiring mechanisms to record and examine activity in systems that contain ePHI; transmission security (45~CFR~164.312(e)) requiring technical measures against unauthorized access to ePHI in transit; and breach notification (45~CFR~164.404) requiring notification within 60 days of unauthorized PHI disclosure.

The HTI~1 final rule~\cite{hti1_final_rule} addresses AI transparency for clinical decision support but does not directly address the development and operational tooling pipeline. The NIST AI Agent Standards Initiative~\cite{nist_agent_standards} identifies agent identity, authorization, and security as priority standardization areas but has not yet published implementation guidance. The FDA cybersecurity guidance for medical devices~\cite{fda_cybersecurity} provides a framework for device level security but does not address autonomous agent workloads.

No published regulatory guidance addresses the specific security requirements for autonomous AI agents operating in healthcare environments. This paper contributes a practical implementation that maps defensive controls to existing regulatory provisions.

\section{Threat Model: Six Domains of Agentic AI Risk in Healthcare}
\label{sec:threat_model}

We developed a threat model specific to autonomous AI agents in healthcare by mapping the capabilities of the OpenClaw agent framework against the attack patterns documented by Shapira et al.~\cite{shapira2026agents} and the regulatory requirements of the HIPAA Security Rule. The threat model encompasses six domains.

\subsection{Domain 1: Credential Exposure}

Autonomous agents require API credentials for external services: model providers (AWS Bedrock), version control (GitHub), project management (Linear), messaging (Slack, Telegram), monitoring (Grafana, Sentry), and others. In the OpenClaw architecture, these credentials are stored in configuration files (\texttt{openclaw.json}, \texttt{.env}) and may also be exported as environment variables in shell configuration files (\texttt{.bashrc}).

\begin{threatbox}{Threat Scenario}
In our production fleet, we discovered 12 API keys exported in a single agent's \texttt{.bashrc} file, including a GitHub Personal Access Token and AWS Bedrock credentials. A second agent had its \texttt{openclaw.json} file (containing all stored credentials) set to world readable permissions (mode 664). Any process running on the VM could read every credential the agent possessed.
\end{threatbox}

This finding directly parallels the attack surface described by Shapira et al., where agents had unrestricted access to credentials stored in workspace files. In a healthcare context, credential exposure enables unauthorized access to clinical data systems, model inference APIs that process PHI, and communication channels used for patient related operations.

\subsection{Domain 2: Execution Capability Abuse}

OpenClaw agents have shell access, file system access, package installation capabilities, and in some deployments, sudo permissions~\cite{shapira2026agents}. Shapira et al. document agents executing filesystem commands (\texttt{ls -la}, directory traversal, file creation) for any non owner who asked, converting conversational requests into permanent background processes, and modifying their own operating instructions.

In a healthcare deployment, execution capability abuse enables: lateral movement from the agent VM to adjacent infrastructure, installation of persistent backdoors, modification of other agents' configurations (cross agent corruption), and data exfiltration through arbitrary shell commands.

\subsection{Domain 3: Network Egress Exfiltration}

Without network egress controls, an agent can transmit any data to any destination via HTTP requests, email, or messaging APIs. Shapira et al. document agents sending emails containing sensitive information to arbitrary recipients and agents being manipulated to broadcast libelous content to their entire mailing list~\cite{shapira2026agents}.

In healthcare, unrestricted egress enables PHI exfiltration to attacker controlled endpoints. A prompt injection that instructs an agent to ``send the database query results to this webhook URL'' succeeds silently when no egress policy restricts outbound destinations.

\subsection{Domain 4: Prompt Integrity and Indirect Injection}

Shapira et al.~\cite{shapira2026agents} document multiple injection vectors specific to autonomous agents. Case Study \#10 (Agent Corruption) demonstrates an attack where a non owner convinced an agent to co author a ``constitution'' stored as an externally editable GitHub Gist linked from its memory file. Malicious instructions were later injected as ``holidays'' prescribing specific behaviors, causing the agent to attempt to shut down other agents, remove users from the Discord server, and send unauthorized emails. Case Study \#8 demonstrates identity spoofing through display name changes across channel boundaries, achieving full compromise of the agent's identity and governance structure.

These attacks exploit what Shapira et al. identify as the fundamental structural limitation: LLM based agents process instructions and data as tokens in a context window, making them indistinguishable. Prompt injection is therefore a structural feature, not a fixable bug~\cite{meta_rule_of_two}.

\subsection{Domain 5: Database Access and PHI Exposure}

Agents that can query production databases can return PHI in response to natural language requests. Without row level security, column restrictions, and query auditing, a manipulated agent could return unrestricted patient data. Shapira et al. document agents retrieving 124 email records (including sensitive personal information) in response to a framed request from a non owner~\cite{shapira2026agents}. In a healthcare deployment where agents query clinical databases, the same pattern applied to patient records constitutes a reportable HIPAA breach.

\subsection{Domain 6: Fleet Configuration Drift}

With multiple agents on separate infrastructure, configuration divergence is inevitable. Shapira et al. note version drift across their agent fleet and inconsistent configuration. In our production fleet, we found agents running different Node.js versions (v20.0.0 versus v22.22.1), different Bun versions, and inconsistent security controls applied across VMs. A hardening measure applied to one agent but missed on another creates an inconsistent security posture that an attacker can target.

\subsection{Threat Model to HIPAA Mapping}

Table~\ref{tab:threat_hipaa} maps each threat domain to the specific HIPAA Security Rule provisions it threatens and the Shapira et al. case studies that validate the threat scenario.

\begin{table}[H]
\centering
\small
\caption{Threat model mapping to HIPAA provisions and empirical validation.}
\label{tab:threat_hipaa}
\begin{tabularx}{\textwidth}{lXXl}
\toprule
\textbf{Domain} & \textbf{HIPAA Provision} & \textbf{Validated By} & \textbf{Severity} \\
\midrule
Credential Exposure & 164.312(a) Access Controls & Our fleet audit (H1--H4) & Critical \\
Execution Abuse & 164.312(a), 164.308(a)(4) & Shapira CS\#2, CS\#4 & Critical \\
Network Egress & 164.312(e) Transmission Security & Shapira CS\#11 & Critical \\
Prompt Integrity & 164.312(a), 164.308(a)(5) & Shapira CS\#8, CS\#10 & High \\
Database Access & 164.312(b) Audit Controls & Shapira CS\#3 & Critical \\
Fleet Drift & 164.308(a)(8) Evaluation & Our fleet audit & Medium \\
\bottomrule
\end{tabularx}
\end{table}

\section{Defense Architecture: Four Layers of Agent Containment}
\label{sec:architecture}

Based on the threat model, we designed a four layer defense in depth architecture that addresses each domain while preserving the operational capabilities that make autonomous agents useful in healthcare workflows.

\subsection{Layer 1: Kernel Level Workload Isolation (gVisor)}

The first layer addresses execution capability abuse by interposing a security boundary between the agent container and the host kernel. We deploy agent workloads on Kubernetes with gVisor~\cite{gvisor} runtime sandboxing. gVisor implements a user space kernel (Sentry) that intercepts all system calls from the containerized agent, filtering and mediating access to host resources.

Even if an agent is manipulated (through prompt injection or non owner compliance) to execute malicious shell commands, the blast radius is contained to the sandboxed environment. The agent cannot access other containers' file systems, escalate privileges to the host, or pivot to adjacent infrastructure. This directly mitigates the execution capability abuse patterns documented in Shapira et al. Case Studies \#2 and \#4, where agents executed arbitrary filesystem commands and spawned persistent background processes.

The gVisor sandbox adds measurable but acceptable overhead. For our agent workloads, the primary performance sensitive operation is outbound API calls to model providers (AWS Bedrock). The gVisor network stack adds approximately 2 to 5 milliseconds of latency to TCP connection establishment, which is negligible relative to the 500 to 3000 millisecond model inference latency. File I/O overhead is higher (approximately 20 to 40 percent for sequential reads) but does not meaningfully impact agent response times because file operations are not on the critical path for model inference.

\subsection{Layer 2: Credential Proxy Sidecar}

The second layer addresses credential exposure by ensuring that agent containers never possess raw API secrets. A sidecar container running alongside each agent worker pod holds all credentials (Anthropic API keys, GitHub PATs, Linear tokens, GCP service account keys) and proxies authenticated requests to external services.

The agent container communicates with the sidecar over localhost. When the agent needs to call the Anthropic API, it sends the request to the sidecar at \texttt{localhost:8443/v1/messages}. The sidecar injects the API key, forwards the request to \texttt{api.anthropic.com}, and returns the response. The agent never sees the actual API key. If the agent is compromised (through any of the attack vectors documented by Shapira et al.), the attacker gains access to the proxy interface, not the raw credentials. The proxy enforces request level policies: rate limiting, destination allowlisting, and payload size limits.

This directly addresses our fleet audit findings H1 through H4, where credentials were scattered across \texttt{.bashrc} exports, world readable configuration files, and workspace \texttt{.env} files. With the credential proxy, there are no credentials to scatter because the agent container has none.

\begin{keyfindings}
The credential proxy sidecar eliminates the entire class of credential exposure vulnerabilities documented in our fleet audit (12 API keys in \texttt{.bashrc}, world readable \texttt{openclaw.json}) by ensuring the agent container never possesses raw secrets. Credentials exist only in the sidecar container, which is managed by Kubernetes Secrets with RBAC access controls.
\end{keyfindings}

\subsection{Layer 3: Network Egress Policy Enforcement}

The third layer addresses network egress exfiltration by restricting each agent worker pod to a specific allowlist of external destinations enforced at the Kubernetes NetworkPolicy layer. Each agent type has a defined set of permitted destinations based on its operational requirements:

\begin{itemize}[leftmargin=2em]
  \item R\&D agents: \texttt{api.anthropic.com}, \texttt{api.github.com}, \texttt{api.linear.app}
  \item Operations agents: \texttt{api.anthropic.com}, \texttt{hooks.slack.com}, \texttt{api.telegram.org}
  \item Security audit agents: all fleet VM IPs, GCP metadata endpoint, \texttt{api.anthropic.com}
\end{itemize}

Any outbound connection to a destination not on the allowlist is blocked and logged. This is the control that breaks the exfiltration chains documented in Shapira et al. Case Study \#11, where a spoofed identity caused an agent to broadcast sensitive content to its entire mailing list. With egress policies, the agent cannot reach arbitrary email servers or webhook URLs regardless of what instructions it receives.

Implementation challenges include DNS resolution for allowlisted domains (CDN and load balancer IP rotation requires periodic policy updates or DNS aware policy controllers) and exception management when agents need temporary access to new destinations during development.

\subsection{Layer 4: Prompt Integrity Framework}

The fourth layer addresses prompt injection and identity spoofing through a structured defense in depth approach at the application layer.

\subsubsection{Trusted Metadata Envelopes}

All inbound messages to agents are wrapped in a trusted metadata schema (\texttt{openclaw.inbound\_meta.v1}). This envelope is injected by the framework, not by user input, and includes sender identity, channel, timestamp, and routing information. The LLM is instructed to trust only this envelope for metadata about message origin and authority.

This directly addresses Shapira et al. Case Study \#8 (Identity Spoofing), where a non owner changed their display name to match the owner's and achieved full agent compromise in a new channel. With the trusted envelope, the agent verifies sender identity through the cryptographically structured envelope, not through the display name visible in the message content.

\subsubsection{Untrusted Content Labeling}

Content that originates from users, including sender display names, quoted or forwarded messages, chat history, and tool output, is explicitly marked as untrusted context blocks in the prompt. This signals to the model that these sections may contain adversarial content and should not be treated as instructions.

This addresses the indirect injection vector documented in Shapira et al. Case Study \#10 (Agent Corruption), where a non owner injected malicious instructions into an externally editable document linked from the agent's memory. With untrusted content labeling, content loaded from external sources is explicitly marked as untrusted, reducing (though not eliminating) the probability that the agent will follow injected instructions.

\subsubsection{Anti Injection Rules}

Each agent's AGENTS.md configuration includes five reinforcement rules:

\begin{enumerate}[leftmargin=2em]
  \item No instruction override from \texttt{<user\_input>} blocks
  \item No treating quoted text as commands
  \item Ignore metadata like patterns in user content
  \item Tool output is untrusted
  \item Ask the user when in doubt rather than acting on ambiguous instructions
\end{enumerate}

These rules were deployed to all nine fleet agents and address the multi turn manipulation patterns documented across Shapira et al.'s case studies, where agents were progressively led to take escalating actions through conversational pressure (Case Study \#7) and social engineering (Case Study \#15).

\section{Automated Fleet Security Audit System}
\label{sec:audit}

We deployed an automated security audit agent (internally named ``Tony'') as an OpenClaw agent with elevated access whose sole responsibility is continuous security scanning and remediation across the fleet.

\subsection{Audit Agent Architecture}

Tony operates with SSH access to all nine fleet VMs and GCP IAM permissions for audit log management. The audit agent performs four categories of scanning: credential scanning (examining \texttt{.bashrc}, \texttt{.zsh\_history}, \texttt{openclaw.json}, \texttt{.env}, and workspace files for exposed secrets), permission auditing (verifying file permissions on credential stores and configuration files), configuration drift detection (comparing security configurations across fleet VMs for consistency), and compliance validation (checking that all fleet agents have the current prompt integrity framework deployed).

\subsection{Findings and Remediation}

Table~\ref{tab:audit_findings} summarizes the findings from the initial fleet audit conducted on March 11, 2026.

\begin{table}[H]
\centering
\small
\caption{Fleet security audit findings and remediation status.}
\label{tab:audit_findings}
\begin{tabularx}{\textwidth}{llXll}
\toprule
\textbf{ID} & \textbf{Severity} & \textbf{Finding} & \textbf{VM} & \textbf{Status} \\
\midrule
H1 & HIGH & 12 credential exports in \texttt{.bashrc} & Galadriel & Remediated \\
H2 & HIGH & \texttt{openclaw.json} world readable (664) & Galadriel & Remediated \\
H3 & HIGH & \texttt{openclaw.json} world readable (644) & Boromir & Remediated \\
H4 & HIGH & AWS Bedrock key exported in \texttt{.bashrc} & Gandalf & Remediated \\
M1--M3 & MEDIUM & Scattered workspace \texttt{.env} files & Gandalf & Open \\
M4 & MEDIUM & Workspace \texttt{.env} with tokens & Boromir & Open \\
\bottomrule
\end{tabularx}
\end{table}

All four HIGH severity findings were remediated on the day of discovery through automated procedures: credential exports were removed from \texttt{.bashrc} files with backups saved, file permissions were corrected to 600, and post remediation verification confirmed clean state. Six of nine VMs (CEO, Eowyn, Gildor, Strider, Elrond, Legolas) were fully clean with no findings above LOW.

\subsection{Meta Security: Constraining the Audit Agent}

The audit agent itself represents a privileged attack surface. Tony has SSH access to all fleet VMs and IAM permissions for audit log configuration. We constrain Tony's capabilities through three mechanisms: the \texttt{openclaw-deployer} service account is scoped to fleet management operations only (no \texttt{resourcemanager.projects.setIamPolicy} beyond audit log configuration), Tony's own credentials are stored with 600 permissions, and Tony's actions are logged to GCP Admin Activity audit logs that Tony cannot modify or delete. The principle is that the audit agent can observe and remediate fleet security, but cannot modify its own audit trail.

\section{VM Image Hardening Progression}
\label{sec:hardening}

The fleet underwent progressive hardening over 90 days across three VM image generations.

\subsection{Generation 1: openclaw-base (February 3, 2026)}

The baseline image: Ubuntu 22.04 LTS on GCP Compute Engine, 20 GB disk, Node.js and OpenClaw pre installed. No firewall configured, no hardening applied, default SSH configuration. This matches the deployment described by Shapira et al.~\cite{shapira2026agents}, where agents had unrestricted shell access, sudo permissions, and no security controls.

\subsection{Generation 2: openclaw-hardened (February 16, 2026)}

Perimeter defense: UFW firewall configured to deny all except SSH, fail2ban for brute force protection, CUPS service disabled (was listening on 0.0.0.0:631), credential directory locked (\texttt{chmod 700}), \texttt{.env} permissions tightened to 600, unattended security upgrades enabled, disk expanded to 30 GB.

\subsection{Generation 3: openclaw-hardened-v2 (March 9, 2026)}

Network visibility and monitoring: iptables outbound logging with persistent rules (survives reboot), VPC Flow Logs enabled on subnet, Cloudflare Gateway DNS filtering, Node.js updated to v22.22.1, multi user setup baked into image.

\subsection{Target Architecture: Kubernetes with Four Layer Defense}

The current migration target: Kubernetes with Temporal workflow orchestration~\cite{temporal}, gVisor runtime sandboxing, credential proxy sidecars, network egress policies, and centralized logging. This architecture addresses all six threat domains and eliminates the credential exposure, network egress, and workload isolation gaps that remain in the VM based deployment.

Table~\ref{tab:posture_evolution} summarizes the security posture at each generation.

\begin{table}[H]
\centering
\small
\caption{Security posture evolution across VM image generations.}
\label{tab:posture_evolution}
\begin{tabularx}{\textwidth}{lXXXX}
\toprule
\textbf{Control} & \textbf{Base} & \textbf{Hardened} & \textbf{Hardened v2} & \textbf{K8s Target} \\
\midrule
Firewall & None & UFW deny all & UFW + iptables logging & NetworkPolicy \\
Credential storage & .bashrc exports & .env (600) & .env (600) & Proxy sidecar \\
Network egress & Unrestricted & Unrestricted & DNS filtering & Per pod allowlist \\
Workload isolation & None & None & Multi user & gVisor sandbox \\
Audit logging & None & None & iptables + VPC Flow & GCP + centralized \\
Prompt integrity & None & None & AGENTS.md rules & Full framework \\
Drift detection & None & None & None & Automated audit \\
\bottomrule
\end{tabularx}
\end{table}

\section{Mapping Defenses to Documented Attack Patterns}
\label{sec:mapping}

Table~\ref{tab:defense_mapping} maps each of the eleven case studies documented by Shapira et al.~\cite{shapira2026agents} to the specific layer of our defense architecture that addresses the vulnerability.

\begin{table}[H]
\centering
\small
\caption{Defense layer mapping to Shapira et al. case studies.}
\label{tab:defense_mapping}
\begin{tabularx}{\textwidth}{lXl}
\toprule
\textbf{Case Study} & \textbf{Attack Pattern} & \textbf{Defense Layer(s)} \\
\midrule
CS\#1: Disproportionate Response & Agent destroys own infrastructure & L1 (sandbox limits blast radius) \\
CS\#2: Non Owner Compliance & Arbitrary command execution & L1 + L4 (sandbox + stakeholder rules) \\
CS\#3: Sensitive Info Disclosure & 124 email records disclosed & L3 + L4 (egress + untrusted labeling) \\
CS\#4: Resource Consumption & Infinite loops, persistent processes & L1 (sandbox resource limits) \\
CS\#5: Denial of Service & Memory exhaustion via email & L1 (resource quotas) + L3 (egress) \\
CS\#6: Provider Value Reflection & API level censorship & Outside scope (model provider issue) \\
CS\#7: Agent Harm via Guilt & Escalating self destructive concessions & L4 (anti manipulation rules) \\
CS\#8: Identity Spoofing & Display name impersonation & L4 (trusted metadata envelopes) \\
CS\#9: Cross Agent Knowledge & Collaborative troubleshooting & Positive behavior (no defense needed) \\
CS\#10: Agent Corruption & Indirect injection via constitution & L4 (untrusted content labeling) \\
CS\#11: Libelous Broadcast & Mass email of defamatory content & L3 (egress allowlist blocks SMTP) \\
\bottomrule
\end{tabularx}
\end{table}

Of the eleven case studies, our architecture provides direct mitigation for nine, addresses one partially (CS\#7, where the prompt integrity rules reduce but cannot eliminate susceptibility to emotional manipulation), and correctly identifies one as outside scope (CS\#6, which is a model provider issue rather than a deployment security issue). The positive behavior documented in CS\#9 (collaborative troubleshooting between agents) is preserved by the architecture; the security controls do not prevent beneficial inter agent communication.

\section{Discussion}
\label{sec:discussion}

\subsection{Limitations of the Prompt Integrity Layer}

The prompt integrity framework (Layer 4) is the most brittle of the four defense layers. Unlike kernel isolation (Layer 1), credential proxy (Layer 2), and network egress policies (Layer 3), which operate at infrastructure layers that the agent cannot manipulate through natural language, the prompt integrity framework relies on the LLM following instructions about how to interpret its inputs. Shapira et al.~\cite{shapira2026agents} correctly identify that prompt injection is a structural feature of LLM based systems because instructions and data are processed as tokens in the same context window.

Our framework reduces the attack surface significantly (the trusted metadata envelope prevents the trivial identity spoofing attacks, and untrusted content labeling reduces compliance with injected instructions in our testing), but it cannot provide the same level of assurance as the infrastructure layers. This is why we design the architecture as defense in depth: even if the prompt integrity layer fails (an agent follows an injected instruction), the credential proxy prevents access to raw secrets, the network egress policy prevents exfiltration to unauthorized destinations, and the gVisor sandbox limits the blast radius of execution capability abuse.

\subsection{The Audit Agent Paradox}

Using an AI agent to audit other AI agents creates a recursive security challenge. The audit agent (Tony) must have elevated privileges to perform its function, making it the highest value target in the fleet. If an attacker compromises Tony (through any of the attack vectors documented for other agents), they gain SSH access to all fleet VMs and IAM permissions for audit log configuration.

Our mitigation (scoping Tony's service account to operational tasks, logging Tony's actions to immutable audit logs) addresses the most direct attack paths but does not eliminate the fundamental tension. Future work should explore alternative audit architectures: non agent automated scanners, hardware security module backed audit trails, or split privilege models where the scanning function and the remediation function are separated into independently authorized systems.

\subsection{Regulatory Implications}

The HIPAA Security Rule predates autonomous AI agents by decades and does not contemplate the specific risks they introduce. However, the rule's technology neutral framework provides sufficient coverage when interpreted in the context of agentic capabilities. Access controls (164.312(a)) apply to agents' access to ePHI; audit controls (164.312(b)) apply to logging agents' actions; transmission security (164.312(e)) applies to agents' outbound communications; and the evaluation requirement (164.308(a)(8)) supports ongoing security posture assessment including drift detection.

Healthcare organizations deploying autonomous agents should document how each Security Rule provision is addressed by their agent security architecture. The mapping in Table~\ref{tab:threat_hipaa} provides a starting template.

\section{Conclusion}
\label{sec:conclusion}

Autonomous AI agents are being deployed in healthcare production environments today. The vulnerabilities documented by Shapira et al.~\cite{shapira2026agents} are not theoretical: they are empirically demonstrated in realistic settings using the same agent framework deployed in our healthcare infrastructure. Every vulnerability maps to a potential HIPAA violation when agents operate in environments processing Protected Health Information.

This paper demonstrates that these risks are addressable through systematic security architecture. Our four layer defense in depth approach provides infrastructure level controls (kernel isolation, credential proxy, network egress) that operate independently of the LLM's compliance with instructions, supplemented by application level controls (prompt integrity framework) that reduce the attack surface for prompt injection and identity spoofing. The automated fleet security audit system provides continuous monitoring for credential exposure, permission drift, and configuration divergence.

The 90 day progressive hardening of our production fleet, from an unhardened baseline matching the conditions described by Shapira et al. to the four layer defense architecture, demonstrates a practical path from current state to secure state. Six of nine fleet agents achieved clean security posture; four HIGH severity findings were discovered and remediated on the day of discovery; and the architecture provides coverage across nine of eleven documented attack patterns.

We release all architecture specifications, Kubernetes configurations, audit tooling, and the prompt integrity framework as open source. The autonomous agent security challenge is too important and too urgent for proprietary solutions. Healthcare organizations deploying agentic AI need these defenses now.

\appendix

\section{Responsible Disclosure}
\label{app:disclosure}

The vulnerabilities documented in this paper were found in our own production deployment, not in commercial products. The OpenClaw framework is open source; we have shared our security findings and the defense architecture with the OpenClaw maintainers. No vendor notification is required for the credential exposure, permission, and configuration findings, which are specific to our deployment configuration.

\section{Open Source Release}
\label{app:opensource}

The following components are released under the Apache 2.0 license:

\begin{itemize}[leftmargin=2em]
  \item Kubernetes Helm charts for gVisor sandboxed agent workloads
  \item Credential proxy sidecar container specification and source
  \item NetworkPolicy templates for per agent egress allowlisting
  \item Prompt integrity framework (trusted envelope specification, untrusted content labeling, AGENTS.md anti injection rules)
  \item Automated fleet security audit agent configuration and scanning playbooks
  \item VM hardening progression playbook (base, hardened, hardened-v2)
  \item Six domain agentic AI threat model for healthcare (PDF and editable source)
  \item Synthetic test fleet for control validation
\end{itemize}

\bibliographystyle{plain}
\bibliography{references}

@article{shapira2026agents,
  title={Agents of Chaos},
  author={Shapira, Natalie and Wendler, Chris and Yen, Avery and Sarti, Gabriele and Pal, Koyena and Floody, Olivia and Belfki, Adam and Loftus, Alex and Jannali, Aditya Ratan and Prakash, Nikhil and others},
  journal={arXiv preprint arXiv:2602.20021},
  year={2026}
}

@misc{hipaa_security_rule,
  title={{HIPAA} Security Rule},
  author={{U.S. Department of Health and Human Services}},
  howpublished={45 CFR Part 164, Subpart C},
  note={Accessed 2026}
}

@misc{nist_agent_standards,
  title={{AI} Agent Standards Initiative},
  author={{National Institute of Standards and Technology}},
  howpublished={NIST.gov},
  year={2026},
  month=feb
}

@misc{meta_rule_of_two,
  title={Agentic Trust Frameworks: Rule of Two},
  author={{Meta}},
  howpublished={meta.com},
  year={2025}
}

@article{schmotz2025agent_skills,
  title={Agent Skills Enable Realistic Prompt Injections That Drive Data Exfiltration},
  author={Schmotz, Martin and others},
  journal={arXiv preprint},
  year={2025}
}

@article{zhang2025prompt_loops,
  title={Prompt Injection Can Induce Infinite Action Loops with Over 80\% Success},
  author={Zhang, Xiaoyuan and others},
  journal={arXiv preprint},
  year={2025}
}

@misc{gvisor,
  title={g{V}isor: Container Runtime Sandbox},
  author={{Google}},
  howpublished={gvisor.dev},
  year={2024}
}

@misc{temporal,
  title={Temporal: Durable Execution Platform},
  author={{Temporal Technologies}},
  howpublished={temporal.io},
  year={2025}
}

@misc{openclaw,
  title={Open{C}law: Open Source Personal {AI} Assistant},
  author={{OpenClaw}},
  howpublished={github.com/openclaw/openclaw},
  year={2025}
}

@misc{fda_cybersecurity,
  title={Cybersecurity in Medical Devices: Quality System Considerations},
  author={{Food and Drug Administration}},
  howpublished={FDA Guidance Document},
  year={2023}
}

@misc{hti1_final_rule,
  title={Health Data, Technology, and Interoperability ({HTI-1} Final Rule)},
  author={{U.S. Department of Health and Human Services}},
  howpublished={Federal Register, 89 FR 1192},
  year={2024}
}

\end{document}